\newif\ifmulticol	\multicoltrue
\newif\ifshowgit	\showgittrue		
\newif\ifgitlocal	\gitlocalfalse		
\newif\ifbiblatex	\biblatexfalse		
\newif\ifbibnum		\bibnumtrue 		
\newif\iflineno		\linenofalse
\newif\iftoc		\tocfalse

\newif\iflucida		\lucidafalse
\newif\ifcm			\cmfalse
\newif\iflibertine	\libertinefalse
\newif\ifcharter	\chartertrue

\newcommand*{\secnumdepth}{0}			
\newcommand*{\tocdepth}{1}				
\newcommand*{\reftitle}{References}		
\newcommand*{\mytitleskip}{-0.2in}		

\multicoltrue\showgittrue\gitlocaltrue\biblatexfalse\bibnumtrue

\newcommand*{\mydocfontsize}{\ifcharter11pt\else\iflibertine11pt\else10pt\fi\fi}
\newcommand*{\setcol}{\ifmulticol twocolumn\else onecolumn\fi}

\documentclass[\mydocfontsize,\setcol]{article}

\newcommand*{\pdfauthor}{Steven A.~Frank}
\newcommand*{\pdftitle}{The common patterns of abundance: the log series and Zipf's law}

\input logseries.sty

\newcommand{\tr}{T}
\newcommand{\trn}{\tr_n}

\newcommand{\trz}{\tr_z}

\newcommand{\nmo}{n^{-1}}

\newcommand*{\Ga}{\alpha}
\newcommand*{\Gb}{\beta}
\newcommand*{\Gd}{\delta}
\newcommand*{\GD}{\Delta}
\newcommand*{\Ge}{\epsilon}
\newcommand*{\Gg}{\gamma}

\newcommand*{\Gl}{\lambda}

\newcommand*{\Gr}{\rho}

\newcommand*{\Gt}{\tau}

\DeclarePairedDelimiter\abs{\lvert}{\rvert}
\DeclarePairedDelimiter\norm{\lVert}{\rVert}
\DeclarePairedDelimiter\angb{\langle}{\rangle}
\DeclarePairedDelimiter\lrb{\lbrack}{\rbrack}
\DeclarePairedDelimiter\lr{\lparen}{\rparen}

\makeatletter
\let\oldabs\abs \def\abs{\@ifstar{\oldabs}{\oldabs*}}
\let\oldnorm\norm \def\norm{\@ifstar{\oldnorm}{\oldnorm*}}
\let\oldangb\angb \def\angb{\@ifstar{\oldangb}{\oldangb*}}
\let\oldlrb\lrb \def\lrb{\@ifstar{\oldlrb}{\oldlrb*}}
\let\oldlr\lr \def\lr{\@ifstar{\oldlr}{\oldlr*}}
\makeatother

\newcommand*{\dd}{\textrm{d}}

\newcommand*{\EEq}[1]{Eqn~\ref{eq:#1}}
\newcommand*{\Eq}[1]{eqn~\ref{eq:#1}}

\newcommand*{\ovr}[2]{{{#1}\over{#2}}}
\newcommand*{\dovr}[2]{\ovr{\dd #1}{\dd #2}}

\newcommand*{\Figure}[1]{Figure~\ref{fig:#1}}

\newcount\BoxNum \BoxNum 1\relax
\makeatletter
\newcommand*{\boxlabel}[1]{%
  \protected@write \@auxout {}{\string \newlabel {box:#1}{{\the\BoxNum}}{}}%
  \advance\BoxNum 1\relax}
\makeatother

\newcommand*{\mytitle}{\pdftitle}

\newcommand*{\myauthor}{Steven A.\ Frank}
\newcommand*{\myaddress}{Department of Ecology and Evolutionary Biology, University of California, Irvine, CA 92697--2525, USA}
\newcommand*{\mycontact}{\href{https://stevefrank.org}{https://stevefrank.org}}

\setabstract{-0.07}{\iftoc-0.5\else0.20\fi}{%
In a language corpus, the probability that a word occurs $n$ times is often proportional to $1/n^2$. Assigning rank, $s$, to words according to their abundance, $\log s$ vs $\log n$ typically has a slope of minus one. That simple Zipf's law pattern also arises in the population sizes of cities, the sizes of corporations, and other patterns of abundance. By contrast, for the abundances of different biological species, the probability of a population of size $n$ is typically proportional to $1/n$, declining exponentially for larger $n$, the log series pattern. This article shows that the differing patterns of Zipf's law and the log series arise as the opposing endpoints of a more general theory. The general theory follows from the generic form of all probability patterns as a consequence of conserved average values and the associated invariances of scale. To understand the common patterns of abundance, the generic form of probability distributions plus the conserved average abundance is sufficient. The general theory includes cases that are between the Zipf and log series endpoints, providing a broad framework for analyzing widely observed abundance patterns. 
}

\begin{document}

\mymaketitle

\iftoc\mytoc{-24pt}{\newpage}\fi

\noindent A few simple patterns recur in nature. Adding up random processes often leads to the bell-shaped normal distribution. Death and other failures typically follow the extreme value distributions. 

Those simple patterns recur under widely varying conditions. Something fundamental must set the relations between pattern and underlying process. To understand the common patterns of nature, we must know what fundamentally constrains the forms that we see. 

Without that general understanding, we will often reach for unnecessarily detailed and complex models of process to explain what is in fact some structural property that influences the invariant form of observed pattern.

We already understand that the central limit theorem explains the widely observed normal distribution \autocite{fischer11a-history}. Similar limit theorems explain why failure often follows the extreme value pattern \autocite{kotz00extreme,coles01an-introduction}. 

The puzzles set by other commonly observed patterns remain unsolved. Each of those puzzles poses a challenge. The solutions will likely broaden our general understanding of what causes pattern. Such insight will help greatly in the big data analyses that play an increasingly important role in modern science.

Zipf's law is one of the great unsolved puzzles of invariant pattern. The frequency of word usage \autocite{zipf35the-psychobiology}, the sizes of cities \autocite{gabaix99zipfs,arshad18zipfs}, and the sizes of corporations \autocite{axtell01zipf} have the same shape. On a log-log plot of rank versus abundance, the slope is minus one. For cities, the largest city would have a rank of one, the second largest city a rank of two, and so on. Abundance is population size. 

The abundance of species is another great unsolved puzzle of invariant pattern. In an ecological community, the probability that a species has a population size of $n$ individuals is proportional to $p^n/n$, the log series pattern \autocite{fisher43the-relation}. Communities differ only in their average population size, described by the parameter, $p$. Actual data vary, but most often fit closely to the log series \autocite{baldridge16an-extensive}.

In this article, I show that Zipf's law and the log series arise as the opposing endpoints of a more general theory. That theory provides insight into the particular puzzles of Zipf's law and species abundances. The analysis also suggests deeper insights that will help to unify understanding of commonly observed patterns.

The argument begins with the invariances that define alternative probability patterns \autocite{frank16common,frank18measurement}. To analyze the invariances of a probability distribution, note that we can write almost any probability distribution, $q_z$, as
\begin{equation}\label{eq:Tz}
  q_z=ke^{-\Gl\trz},
\end{equation}
in which $\tr(z)\equiv\trz$ is a function of the variable, $z$. The probability pattern, $q_z$, is invariant to a constant shift, $\trz\mapsto a+\trz$, because we can write the transformed probability pattern in \Eq{Tz} as
\begin{equation*}
  q_z=k_ae^{-\Gl\lr{a+\trz}}=ke^{-\Gl\trz},
\end{equation*}
with $k=k_ae^{-\Gl a}$. We express $k$ in this way because $k$ adjusts to satisfy the constraint that the total probability be one. In other words, conserved total probability implies that the probability pattern is shift invariant with respect to $\trz$.

Now consider the consequences if the average of some value over the distribution $q_z$ is conserved. That constraint causes the probability pattern to be invariant to a multiplicative stretching (or shrinking), $\trz\mapsto b\trz$, because
\begin{equation*}
  q_z=ke^{-\Gl_b b\trz}=ke^{-\Gl\trz},
\end{equation*}
with $\Gl=\Gl_b b$. We specify $\Gl$ in this way because $\Gl$ adjusts to satisfy the constraint of conserved average value. Thus, invariant average value implies that the probability pattern is stretch invariant with respect to $\trz$.

Conserved total probability and conserved average value cause the probability pattern to be invariant to an affine transformation of the $\trz$ scale, $\trz\mapsto a+b\trz$, in which ``affine'' means both shift and stretch.

The affine invariance of probability patterns with respect to $\trz$ induces significant structure on the form of $\trz$ and the associated form of probability patterns. Understanding that structure provides insight into probability patterns and the processes that generate them  \autocite{frank11a-simple,frank14how-to-read,frank16common}.

In particular, Frank \& Smith \autocite{frank11a-simple} showed that the invariance of probability patterns to affine transformation, $\trz\mapsto a+b\trz$, implies that $\trz$ satisfies the differential equation
\begin{equation*}
  \dovr{\trz}{w}=\Ga+\Gb\trz,
\end{equation*}
in which $w(z)$ is a function of the variable $z$. The solution of this differential equation expresses the scaling of probability patterns in the generic form
\begin{equation}\label{eq:wmetric}
  \trz\stackrel{}{=}\frac{1}{\Gb}\lr{e^{\Gb w}-1},
\end{equation}
in which, because of the affine invariance of $\trz$, I have added and multiplied by constants to obtain a convenient form, with $\trz\rightarrow w$ as $\Gb\rightarrow0$. With this expression for $\trz$, we may write probability patterns generically as
\begin{equation}\label{eq:canonical}
  q_z=ke^{-\Gl \lr{e^{\Gb w}-1}/\Gb}.
\end{equation}

Turning now to the log series and Zipf's law, the relation $n=e^r$ between observed pattern, $n$, and process, $r$, plays a central role. Here, $r$ represents the total of all proportional processes acting on abundance. A proportional process simply means that the number of individuals or entities affected by the process increases in proportion to the number currently present, $n$.

The sum of all of the proportional processes acting on abundance over some period of time is
\begin{equation*}
  r = \int_0^\Gt m(t)\dd t.
\end{equation*}
Here, $m(t)$ is a proportional process acting at time $t$ to change abundance. The value of $r=\log n$ is the total of the $m$ values over the total time, $\Gt$. For simplicity, I assume $n_0=1$.

Proportional processes are often discussed in terms of population growth \autocite{gibrat31les-inegalites,gabaix99zipfs}. However, many different processes act individually on the members of a population. If the number of individuals affected increases in proportion to population size, then the process is a proportional process. 

Growth and other proportional processes often lead to an approximate power law, $q_n\approx kn^{-\Gr}$. However, the exponent of a growth process does not necessarily match the values observed in the log series and Zipf's law. We need both the power law aspect of proportional process and something further to get the specific forms of those widely observed abundance distributions. That something further arises from conserved quantities and their associated invariances.

The log series and Zipf's law follow as special cases of the generic probability pattern in \Eq{canonical}. To analyze abundance, focus on the process scale by letting the variable of interest be $z\equiv r$, with the key scaling simply the process variable itself, $w(r)=r$. Then \Eq{canonical} becomes
\begin{equation}\label{eq:canonicalR}
  q_r\dd r=ke^{-\Gl \lr{e^{\Gb r}-1}/\Gb}\,\dd r,
\end{equation}
in which $q_r\dd r$ is the probability of a process value, $r$, in the interval $r+\dd r$. From the relation between abundance and process, $n=e^r$, we can change from the process scale to the abundance scale by the substitutions $r\mapsto\log n$ and $\dd r\mapsto \nmo\dd n$, yielding the identical probability pattern expressed on the abundance scale
\begin{equation}\label{eq:canonicalN}
  q_n\dd n=k\nmo e^{-\Gl \lr{n^\Gb-1}/\Gb}\,\dd n.
\end{equation}
The value of $k$ always adjusts to satisfy the constraint of invariant total probability, and the value of $\Gl$ always adjusts to satisfy the constraint of invariant average value. 

For $\Gb=1$, we obtain the log series distribution
\begin{equation}\label{eq:logseriesN}
  q_n=k\nmo e^{-\Gl n},
\end{equation}
replacing $n-1$ by $n$ in the exponential term which, because of affine invariance, describe the same probability pattern. The log series is often written with $e^{-\Gl}=p$, and thus $q_n=kp^n/n$. One typically observes discrete values $n=1,2,\dots$. The Appendix shows the relation between discrete and continuous distributions and the domain of the variables \autocite{au99transforming}. The continuous analysis here is sufficient to understand pattern.  

For $\Gb\rightarrow0$, we have $\lr{n^\Gb-1}/\Gb\rightarrow\log n$, which yields
\begin{equation}\label{eq:zipfN}
  q_n=\Gl n^{-(1+\Gl)}
\end{equation}
for $n\ge1$. If we constrain average abundance, $\angb{n}$, with respect to this distribution, then 
\begin{equation*}
  \Gl=\frac{1}{1-1/\angb{n}}.
\end{equation*}
For any average abundance that is finite and not small, $\Gl\rightarrow1$, which is Zipf's law.  

\EEq{canonicalN} provides a general expression for abundance distributions. The log series and Zipf's law set the endpoints of $\Gb=1$ and $\Gb\rightarrow0$. We can understand the differences between abundance distributions in terms of the parameter $\Gb$ by writing the distribution in the generic form of \Eq{Tz}, with the defining affine invariant scale
\begin{equation}\label{eq:Tn}
  \trn= \frac{\log n}{\Gl} + \frac{n^\Gb-1}{\Gb}.
\end{equation}
This scale expresses the invariances that define the pattern. At the Zipf's law endpoint, $\Gb\rightarrow0$, the scale becomes $2\log n=2r$, when satisfying the constraint that the average abundance, $\angb{n}$, is sufficiently large. 

In this case, with affine invariant scale $\trn=2r$, neither addition nor multiplication of process value, $r\mapsto a+br$, alters the pattern. We could have started with this affine invariance, and derived the probability pattern from the invariance properties \autocite{frank16common,frank18measurement}. 

For the log series endpoint, $\Gb=1$, the affine invariant scale is
\begin{equation*}
  \trn=\frac{1}{\Gl}\log n + n.
\end{equation*}
The dominant aspect of the scale changes with $n$. For small abundances, the logarithmic scale $r=\log n$ dominates, and for large abundances, the linear scale $n=e^r$ dominates. Many common probability patterns change their scaling with magnitude \autocite{frank14how-to-read,frank16the-invariances}.

For log series patterns, the dominance of scale at small magnitude by $r$ corresponds to affine invariance with respect to $r$. At larger abundances, the dominance by the effectively linear scale, $n$, corresponds to invariance to a shift in process $r\mapsto a+r$, but not to a multiplication of process, $r\mapsto br$, because $e^{br}=n^b$ is a power transformation of abundance. Linear scales are not invariant to power transformations. Once again, we could have derived the pattern from the invariances.

In \Eq{Tn}, intermediate values of $\Gb$ combine aspects of Zipf's law and the log series. The closer $\Gb$ is to one of the endpoints, the more the invariance characteristics of that endpoint dominate pattern. 

This analysis shows how two great and seemingly unconnected puzzles solve very simply in terms of a single continuum between alternative invariances. This approach reveals the simple invariant structure of many common probability patterns.

\section*{Acknowledgments}

\noindent The Donald Bren Foundation supports my research. I completed this work while on sabbatical in the Theoretical Biology group of the Institute for Integrative Biology at ETH Zürich.


\mybiblio	

\addcontentsline{toc}{section}{Appendix: Discrete and continuous distributions}

\newpage
\section*{Appendix: Discrete and continuous distributions}

Discrete and continuous probability distributions are usually analyzed differently, which prevents a general understanding of scale. This Appendix presents a method by which a change of variable or a change of scale can be done in a consistent way for both discrete and continuous distributions. The final section relates discrete and continuous scales, illustrated by the log series.

For example, suppose initial measurements are in terms of abundance, $n$, and we wish to analyze the data on the transformed logarithmic scale, $r=\log n$. How can we make the change of variable, $n\mapsto e^r$, consistently for discrete and continuous cases? 

The Dirac delta function provides the basis for a consistent method. The next section introduces the basic aspects and notation for the Dirac delta function. The following section shows how to use this method to obtain a consistent approach for transforming scale by change of variable. The final sections consider transformations between discrete and continuous variables and the specification of the domains of variables.

\subsection{Dirac delta function}

The Dirac delta function, $\Gd$, provides the key. The function is defined such that 
\begin{equation*}
  \int_{-\Ge}^{\Ge} \Gd(z)\dd z=1
\end{equation*}
for any real value $\Ge>0$. In other words, for any region of integration containing 0, the integral of $\Gd(z)$ is one. Then we also have
\begin{equation*}
  \int_{z-\Ge}^{z+\Ge} f(x)\Gd(x-z)\dd x=f(z).
\end{equation*}
In other words, the integral picks out the function evaluated at the point $x=z$, at which $\Gd(x-z)=\Gd(0)$.  

With that definition, we can write a discrete probability distribution at the set of points $\Omega=\left\{x_i\right\}$ as a continuous probability density function
\begin{equation}\label{eq:discrete1}
  f(x)\sum_{x_i\in\Omega}\Gd(x-x_i)=f(x)\Gd_x,
\end{equation}
because the cumulative distribution function, $F(x)$, of the continuous density, $f(x)\Gd_x$, has the form of a discrete probability distribution
\begin{equation*}
  F(a)=\int_{-\infty}^{a+\Ge} f(x)\Gd_x\dd x=\sum_{x_i\in\Omega} f(x_i),
\end{equation*}
in which $x_i<a+\Ge$ for an infinitesimal positive value, $\Ge$. We need the extra $\Ge$ so that $\Gd_x$ integrates to one around a point $x_i=a$, 

For continuous distributions, let the density of points in $\Omega$ increase to fill the interval $(-\infty,\infty)$ continuously. Then
\begin{equation*}
  \Gd_x=\sum_{x_i\in\Omega}\Gd(x-x_i)\rightarrow\int_{-\infty}^\infty \Gd(x-x_i)\dd x_i=1,
\end{equation*}
because, whatever the value of $x$, there will be some point $x_i\in\Omega$ for which $x=x_i$, and any integral over the region including that point is one. With $\Gd_x=1$, the continuous probability density function is $f(x)$, and the cumulative distribution function is
\begin{equation*}
  F(a)=\int_{-\infty}^{a+\Ge} f(x)\dd x=\int_{-\infty}^{a} f(x)\dd x,
\end{equation*}
because $\int_{a}^{a+\Ge} f(x)\dd x=0$ for infinitesimal $\Ge$ and finite $f(x)$. 

\subsection{Change of variable}

I seek a consistent method for doing a change of variable in both continuous and discrete probability distributions. The approach arises from always considering the probability associated with a value as the area of a rectangle. 

For a continuous distributions, we write $f(x)\dd x$, which is the product of the probability function, $f$, as the height, and the infinitesimal interval, $\dd x$, as the width. Thus, the probability in the interval $(a,a+\Ge)$, with small width $\Ge$, is
\begin{equation*}
  \int^{a+\Ge}_a f(x)\dd x\approx f(a)\Ge,
\end{equation*}
the product of the height, $f(a)$, and the width, $\Ge$. When we change variables, $x\mapsto g(x)\equiv y$, we obtain both a new height, $f(y)$, and a new width, $\dd y$, and so we must compensate appropriately, as shown below.

For discrete distributions, we may write $f(x)\GD x$, in which $\GD x$ is the width associated with a discrete point, $x$. Typically, we assume that $\GD x=1$ for all $x$, and write the discrete probability as $f(x)$. We can think of this as the area of a rectangle with implicit width of one. 

When we change variables, $x\mapsto g(x)\equiv y$, traditionally one keeps the interval widths, $\GD y=1$, as one on the new scale, $y$, and the probabilities are simply $f(y)$ at the new points, $y$. However, by changing scales, the spacing between the discrete points on the $y$ scale differs from the spacing between points on the original $x$ scale. 

This change of spacing can be interpreted as a change in the widths associated with discrete probability points, or as a change in the density of probability points in intervals along the $y$ scale. Thus, as in the continuous case, we may wish to keep track of how both the heights change, $f(x)\mapsto f(y)$, and how the widths change with a change of scale, $\GD x\mapsto \GD y$. Doing so provides a consistent way of changing variables for continuous and discrete cases.

I begin with the standard method for continuous variables. I then demonstrate an analogous method for discrete variables based on the Dirac delta function.

For a continuous distribution, $f(x)$, I make the change $x\mapsto g(x)\equiv y$. With that transformation, we have
\begin{equation*}
  \dd y/\dd x=g'(x).
\end{equation*}
Define
\begin{equation*}
  m_y = \frac{1}{\abs{g'(x)}},
\end{equation*}
in which the absolute value arises because we are using $\dd x$ and $\dd y$ as positive probability measures. Thus,
\begin{equation*}
  \dd x = m_y\dd y.
\end{equation*}
Then the standard result for the change of variable $x\mapsto y$ in a continuous distribution yields
\begin{equation}\label{eq:contChange}
  f(x)\dd x = f(y)m_y\dd y.
\end{equation}

For discrete distributions, I will derive the analogous change of variable expression
\begin{equation}\label{eq:discreteChange}
  f(x)\Gd_x\dd x = f(y)m_y\Gd_y'\dd y=f(y)\Gd_y\dd y.
\end{equation}
To obtain this result, we need to show that the change of variable $x\mapsto g(x)\equiv y$ leads to
\begin{equation*}
  \Gd_x\dd x\mapsto m_y\Gd_y'\dd y=\Gd_y\dd y,
\end{equation*}
which follows if 
\begin{equation*}
  \Gd_x\mapsto m_y^{-1}\Gd_y\equiv\Gd_y'.
\end{equation*}
To obtain this expression for $\Gd_y'$, we need the general change of variable rule for the Dirac delta function
\begin{align*}
  \Gd(x-x_i)&\mapsto\abs{g'(x_i)}\Gd\lrb{g(x)-g(x_i)}\\[4pt]
  			&=m_y^{-1}\Gd(y-y_i).
\end{align*}
Thus, with $\Omega'=\left\{g(x_i)\right\}=\left\{y_i\right\}$, we have
\begin{equation*}
  \Gd_x=\sum_{x_i\in\Omega}\Gd(x-x_i)\mapsto
  	m_y^{-1}\sum_{y_i\in\Omega'}\Gd(y-y_i)=m_y^{-1}\Gd_y.
\end{equation*}

\subsection{The gamma and log series distributions}

The gamma distribution is given by the probability function
\begin{equation*}
  f(x)=kx^{\Ga-1}e^{-\Gl x},
\end{equation*}
in which the constant $k$ normalizes the total probability to be one. With $\Ga=0$ and $x>x_0>0$ for $x_0$ not too close to zero, this has the same mathematical form as the log series distribution. 

Consider the change in variable $r=\log x=g(x)$, which corresponds to $x\mapsto e^r$. Then,
\begin{equation*}
  \abs{g'(x)}=\dd\log x/\dd x=1/x=e^{-r}=m_r^{-1}.
\end{equation*}
If we consider $f(x)$ as a continuous distribution, then we can apply the formula for change of variable in \Eq{contChange} to obtain
\begin{align*}
  f(x)\dd x &= f(r)m_r\dd r\\[4pt]
  			&=ke^{r(\Ga-1)}e^{-\Gl e^r}e^r\dd r\\[4pt]
        	&=ke^{r\Ga-\Gl e^r}\dd r\\[4pt]
        	&=h(r)\dd r,
\end{align*}
in which
\begin{equation}\label{eq:hr}
  h(r)=f(r)m_r=ke^{r\Ga-\Gl e^r},
\end{equation}
for $r>\log x_0$. For $\Ga=0$, transforming the log series form 
\begin{equation*}
  f(x)=kx^{-1}e^{-\Gl x}
\end{equation*}
by $r=\log x$ yields the equivalent distribution on the $r$ scale as
\begin{equation}\label{eq:hrlogseries}
  h(r) =ke^{-\Gl e^r}.
\end{equation}
Now consider $f(x)$ as a discrete distribution. Then, by \Eq{discreteChange}, we immediately have
\begin{equation*}
   f(x)\Gd_x\dd x = f(r)m_r\Gd_r'\dd r=h(r)\Gd_r'\dd r,
\end{equation*}
in which the full form of $h(r)$ is given in \Eq{hr}, and we also have the log series form with $\Ga=0$ in \Eq{hrlogseries}. 

Thus, by using the measure $\dd r$ for the widths in the continuous case and the measure $\Gd_r'\dd r$ for the widths in the discrete case, we obtain the identical probability function $h(r)$ for the continuous and discrete cases. 

\begin{figure*}[t]
\centering
\includegraphics[width=\dimexpr \hsize - \dimexpr 0.5in\relax]{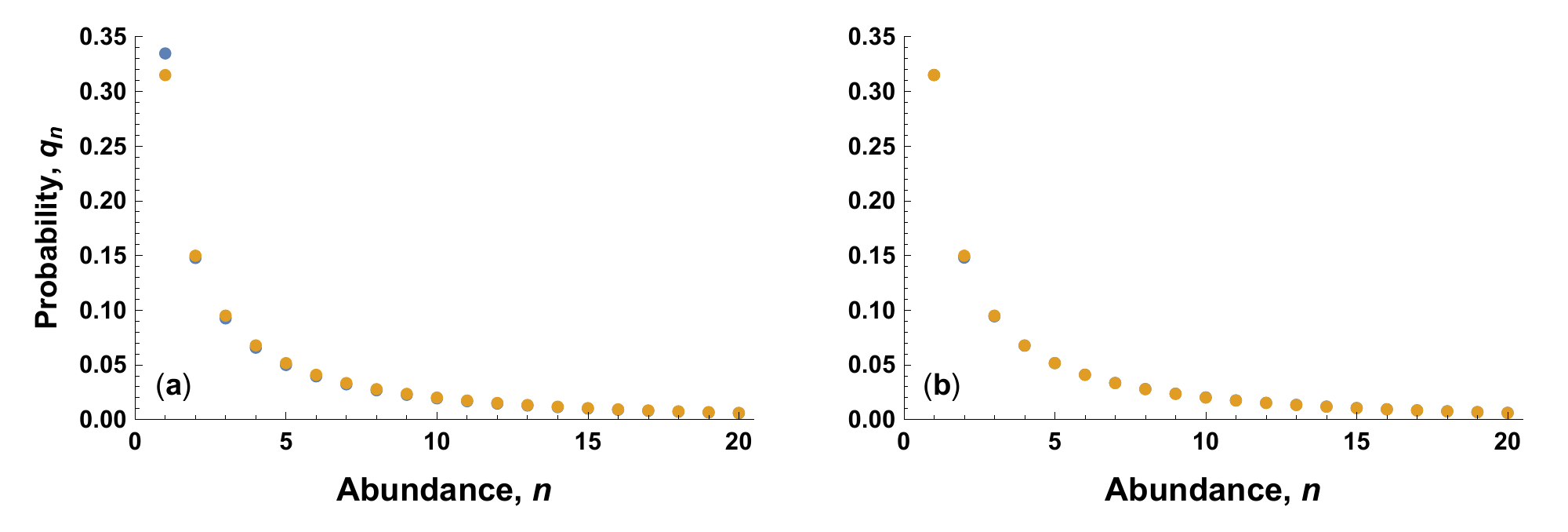}
\caption{Close match between the continuous distribution for process, $r$, and the discrete log series for abundance, $n$. The blue circles show the probabilities of the discrete values of $n$ obtained from the continuous distribution in $r$, calculated by \Eq{cont2disc}. The gold circles show the actual values of $q_n$ for the log series in \Eq{apxqn}. For most points, the values are nearly identical, causing the gold circles to hide the underlying blue circles. (a) When using no offset for continuous intervals, $\Gg=0$, a slight mismatch occurs, particularly at $n=1$. The nonlinearity of $q_r$ causes the mismatch. (b) When using an offset of $\Gg=0.1$, the continuous distribution of the process, $r$, maps almost perfectly onto the discrete log series of abundance, $n$. For all calculations, $\Gl=0.05$. }
\label{fig:cont2discr}
\end{figure*}

For the discrete case, we must keep in mind that
\begin{equation*}
  h(r)\Gd_r'\dd r=h(r)m_r^{-1}\Gd_r\dd r,
\end{equation*}
in which the right side is the traditional expression that picks out the probability mass, $h(r)m_r^{-1}$, as the heights at the points defined by $\Gd_r\dd r$, implicitly using constant widths of one on all scales, because at a discrete point, $r^*$, at which the probability is nonzero, 
\begin{equation*}
  \int_{r^*-\Ge}^{r^*+\Ge}\Gd_r\dd r=1.
\end{equation*}

In the gamma example, $m_r^{-1}=e^{-r}$, with $x=1,2,\dots$ and $r=\log 1, \log 2, \dots$, we have the traditional expression for a discrete change of variable with constant widths of one as
\begin{align*}
  f(x)\Gd_x\dd x&\mapsto h(r)m_r^{-1}\Gd_r\dd r\\[5pt]
  	&=ke^{r(\Ga-1)-\Gl e^r}\Gd_r\dd r.
\end{align*}
For the log series case, $\Ga=0$, this becomes
\begin{equation*}
  h(r)m_r^{-1}\Gd_r\dd r=ke^{-r-\Gl e^r}\Gd_r\dd r.
\end{equation*}
We can go back to the classic log series expression by reversing the change, $e^r\mapsto n$, yielding the discrete distribution $f(n)\Gd_n\dd n$ for $n=1,2,\dots$, with
\begin{equation*}
  f(n)=k\nmo e^{-\Gl n}.
\end{equation*}

In these examples, I have assumed that the distributions on the $r$ and $n\equiv x$ scales are either both discrete or both continuous. In application, it will usually make sense to think of process, $r$, as a continuous variable, and abundance, $n$, as a discrete variable. Therefore, we need to consider transformations between continuous and discrete variables. I discuss that topic in the final section, after a brief summary of the discrete transformations.

\subsection{Summary of alternative discrete expressions}

We have two different ways of expressing transformed discrete variables, in which the initial distribution is given by $f(x)\Gd_x\dd x$, and we transform $x\mapsto y$. 

In the first expression, the transformed distribution is 
\begin{equation*}
  f(y)m_y\Gd_y'\dd y=h(y)\Gd_y'\dd y,
\end{equation*}
in which $h(y)=f(y)m_y$ is the same expression as obtained when transforming continuous variables. The measure for the $y$ scale, $\Gd_y'\dd y$, stretches or shrinks in relation to the measure for the $x$ scale, altering the widths associated with each height. This form has the advantage of retaining the same expressions for the probability functions in the discrete and continuous cases.

In the second expression, the transformed distribution is 
\begin{equation*}
  f(y)\Gd_y\dd y=h(y)m_y^{-1}\Gd_y\dd y,
\end{equation*}
in which this expression highlights the difference between the standard form of the probability function obtained in the discrete case, $f(y) = h(y)m_y^{-1}$, and the standard form of the probability function obtained in the continuous case, $h(y)=f(y)m_y$. Here, we assume the widths associated with each probability point remain one on all scales. 

For the transformation $x\mapsto g(x)\equiv y$, the value $m_y^{-1}=\abs{g'(x)}$ determines the distinction between the discrete and continuous cases, associated with the change in widths between scales.

\subsection{Transforming between continuous and discrete variables}

In the prior cases, we transformed from one discrete variable to another discrete variable or from one continuous variable to another continuous variable. The expressions for transformation followed without any further assumptions.

In the case of the log series and the distribution of abundances, it often make sense to consider process, $r$, as a continuous variable, and abundance, $n$, as a discrete variable. We usually think of process as causing abundance. So we should begin with the continuous distribution for process, $q_r$, and seek the corresponding discrete distribution for abundance, $q_n$.

The probability mass, $q_n$, at a particular value of $n=1,2,\dots$, should map to the total probability for a matching range of growth rates, such that
\begin{equation}\label{eq:cont2disc}
  q_n=\int_{a_n}^{b_n} q_r\dd r,
\end{equation}
in which $r>a_1$. 

We need the particular form of $q_r$, which we take as the fundamental shift-invariant distribution in the main text
\begin{equation}\label{eq:apxqr}
  q_r=ke^{-\Gl e^r}.
\end{equation}
We also need, for each $n$, the interval of growth rates, $(a_n,b_n)$, that maps to the abundance, $n$. In particular, we need a sequence of contiguous intervals, $\left\{\lr{a_n,b_n}\right\}$, that associate each value of $n$ to an interval for $r$, such that $a_{n+1}=b_n$. 

The problem concerns how to pick the sequence of intervals. The simplest approach is to use a standard rounding procedure, such that 
\begin{align*}
  a_n&=\log(n-0.5+\Gg)\\
  b_n&=\log(n+0.5+\Gg),
\end{align*}
in which $\Gg<0.5$ is a correction for centering intervals to account for the nonlinearity in the mapping between the continuous and discrete probability expressions.

If we use $\Gg=0.1$, the transformation from the continuous scale $r$ to the discrete scale $n$ in \Eq{cont2disc} yields a distribution that closely matches the log series
\begin{equation}\label{eq:apxqn}
   q_n\approx kn^{-1}e^{-\Gl n}.
\end{equation}
\Figure{cont2discr} shows the match between the continuous distribution for $r$ and the associated discrete distribution for $n$. 


\end{document}